# MULTI-ALGORITHM ANALYSIS OF THE SEMI-REGULAR VARIABLE DY PER, THE PROTOTYPE OF THE CLASS OF COOL RCRB VARIABLES

Ivan L. Andronov[1], Kateryna D. Andrych[1,2], Lidia L. Chinarova[1,3]
[1] Department of Mathematics, Physics and Astronomy
Odessa National Maritime University, Odessa, Ukraine
[2] Department of Theoretical Physics and Astronomy
Odessa I.I.Mechnikov National University, Odessa, Ukraine
[3] Astronomical Observatory
Odessa I.I.Mechnikov National University, Odessa, Ukraine

*tt_ari@ukr.net, katyaandrich@gmail.com, llchinarova@gmail.com*

Abstract. Multiple algorithms of time series analysis are briefly reviewed and partially illustrated by application to the visual observations of the semi-regular variable DY Per from the AFOEV database. These algorithms were implemented in the software MCV (Andronov and Baklanov, 2004), MAVKA (Andrych and Andronov, 2019; Andrych et al., 2019). Contrary to the methods of "physical" modeling, which need to use too many parameters, many of which may not be determined from pure photometry (like temperature/spectral class, radial velocities, mass ratio), "phenomenological" algorithms use smaller number of parameters. Beyond the classical algebraic polynomials, in the software MAVKA are implemented other algorithms, totally 21 approximations from 11 classes.

Photometric observations of DY Per from the AFOEV international database were analyzed. The photometric period has switched from $P=851.1^d\pm4.1^d$ to $P=780.5^d\pm2.7^d$ after JD $2454187\pm9^d$.

A parameter of sinusoidality is introduced, which is equal to the ratio of effective semi-amplitudes of the signal determined from a sine fit and the running parabola scalegram.

**Introduction**

DY Per is classified in the "General Catalogue of Variable Stars" [1,2] as an SRb-type pulsating variable with a period of $900^d$ and a range of brightness variations $10.6^m$-$13.2^m$ (V) and spectral class C4,5(R8). No period or initial epoch is mentioned in the "Variable Stars Index" (VSX) [3]. It was suspected to be an R CrB star based on the photometry by Alksnis [4]. He re-estimated a period to be $792^d$ and reported on irregular decline events (weakenings) occurring with an interval $726^d$–$934^d$, with a mean value ~810d, close to the main period. Alcock et al. [5] separated the stars similar to DY Per, to a separate class. They resemble R CrB, but have much lower temperatures ~3500K. Tisserand et al. [6] discussed the connection between RCrBs, DYPers, and ordinary carbon stars. So these stars are phenomenologically SRb, but are intermediate between them and R CrB. The R CrB stars are explained as a result of a merger of white dwarf companions in a binary system. Detailed classification of variable stars is listed in the GCVS [1,2] and numerous monographs (e.g. [7-9])

The periodogram analysis (using the least squares sine approximation [10,11]=TP1=trigonometric polynomial of order 1) showed the periods of $857^d \pm 3^d$, $367^d \pm 1^d$ and $618^d \pm 4^d$ with semi-amplitudes $R_{TP}$ of $1.13^m$, $0.39^m$ and $0.30^m$, respectively. The periods from the wavelet periodogram (using the improved modification [12,13]) has shown two similar values $857^d$, $366^d \pm 1^d$ and a very different $248^d$ [14].

In this paper, we analyze visual observations from the AFOEV database for the time interval following that studied in the catalogue by Chinarova and Andronov [14]. Totally, after filtration of unsure (:), fainter than (<) data and outliers [15] , there remained n=3762 data points obtained on JD 2451633–2458027, the range of magnitudes $10.4^m$-$16.32^m$. we apply different methods for data analysis.

**Periodogram Analysis**

We have applied the least squares sine approximation [10,11] implemented in the software MCV [16]. The periodogram is shown in Fig. 1. There are some peaks, the highest of which corresponds to the period $P=1/f=794.6^d \pm 1.0^d$, which is shorter than the GCVS [1] value of 900d and more recent value $857^d \pm 3^d$ [14]. This may argue for a possible period decrease during the period of observations, which will be checked below. The initial epoch $T_0=2454431.4 \pm 2.5^d$, semi-amplitude $0.88^m \pm 0.02^m$ and a mean (over the period) value $11.780^m \pm 0.013^m$.

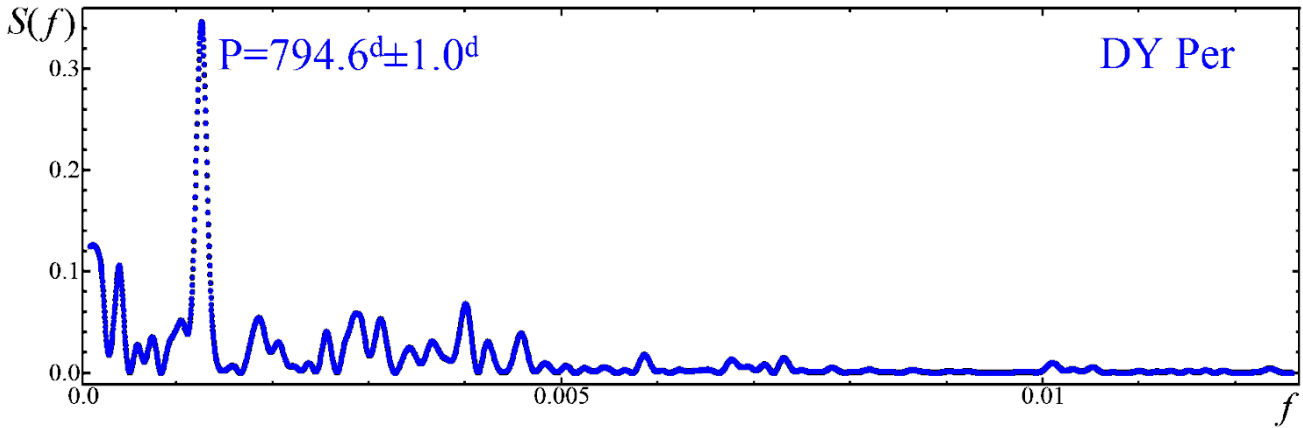

Fig. 1. Periodogram $S(f)$ [10] of DY Per. The highest peak corresponds to the period $P=1/f=794.6^d \pm 1.0^d$.

The light curve is shown in Fig. 2. It is clearly visible that there are systematic deviations of the light curve from a mentioned sine curve either in the shape, or in the pulsation-averaged brightness.

Below, we describe various algoritms. For suitable comparison, they are all shown in a single Figure 2. The legends are: TP1 – Trigonometric Polynomial of the first order (sime); RP – Runnimg Parabola; P – Polynomial; AP – asymptotic parabola; PS – parabolic spline; SP – symmetrical polynomial; NAV* – modified "New Algol Variable", RS – running sine.

The second peak (in height) corresponds to a long-term "period" $7543^d \pm 543^d$, semi-amplitude $0.544^m \pm 0.024^m$, $T_0=2451398^d \pm 47^d$. However, this "period" exceeds the duration of observations, thus may be correctly named as a "possible cycle" or "characteristic time-scale".

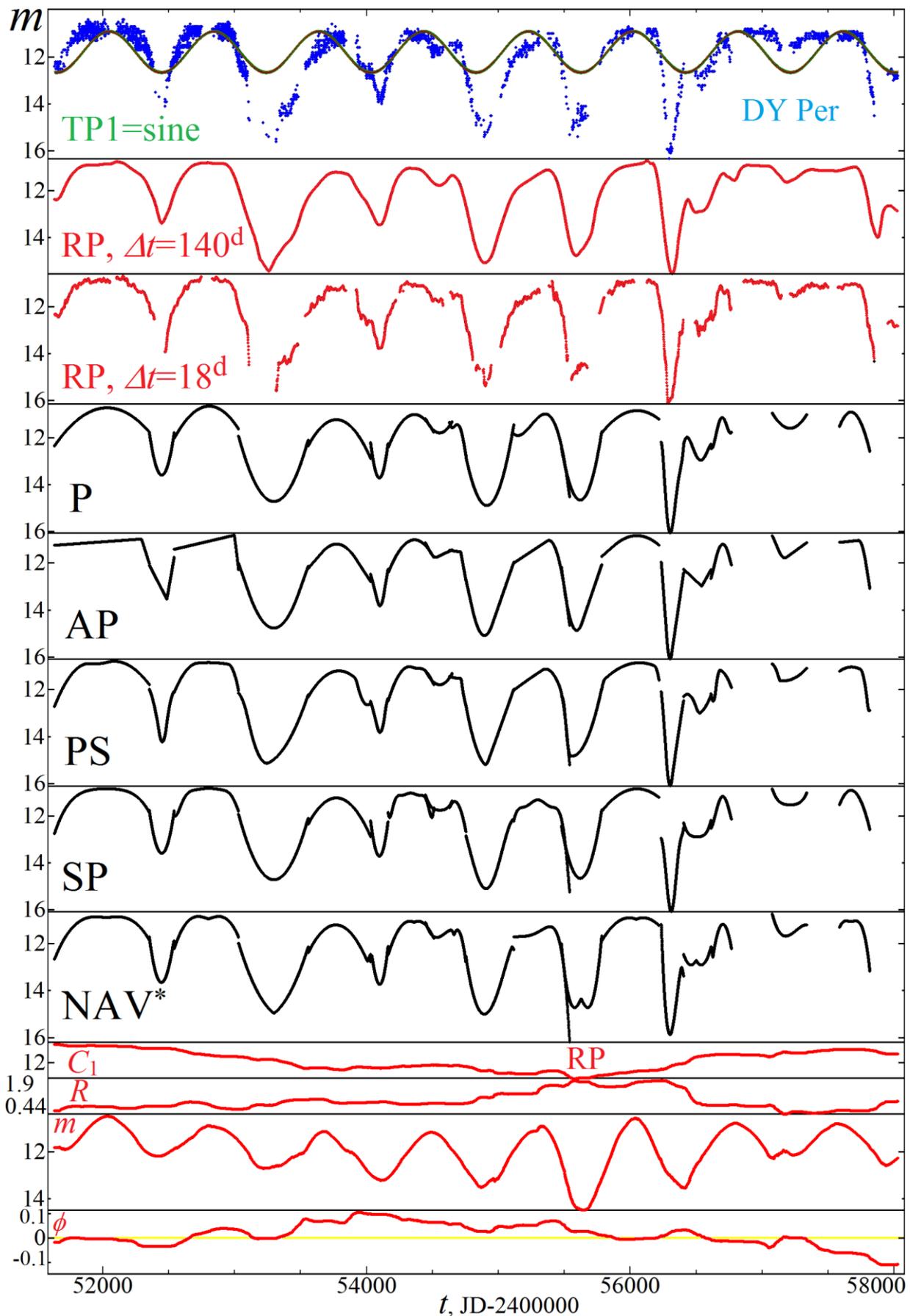

Fig. 2. Light curve of DY Per from the AFOEV database and its approximations using

various algorithms. In the bottom RS block, there are some more dependencies, i.e. the semi-amplitude $R$ and the mean over the period $C_1$, the approximation $m$ and phase $\phi$.

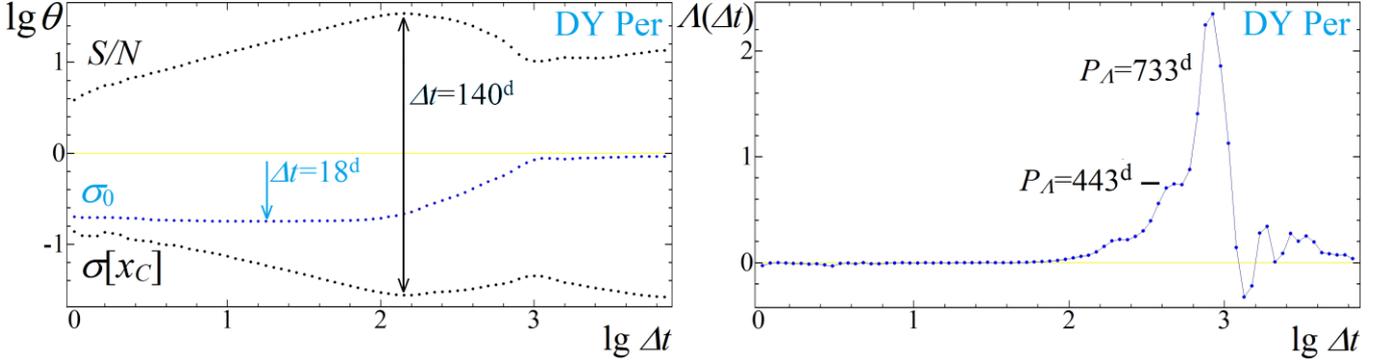

Fig 3. Left: scalegrams using the Running Parabola (RP) algorithm: Vertical lines correspond to minimum of $\sigma_0$ ($18^d$), minimum of $\sigma[x_C]$ ($140^d$), and maximum of $S/N$ ($140^d$). Right: the "Lambda" $\Lambda$–scalegram.

### Scalegram Analysis Using Running Parabolae

Andronov [17] introduced a complete set of equations describing the "running approximations" with arbitrary basic and weight functions, improved for the arguments, which are (generally) irregular. This is a typical case for astronomical observations from space and ground-based surveys.

For the signals with very low coherence, with drastic variations of the individual oscillations, the "running parabola" was proposed with the weight function $p(z)=(1-z^2)^2$, $z=(t-t_0)/\Delta t$, $t$ is the time of observation, $t_0$ is time of the center of interval of smoothing, and $\Delta t$ is the filter half-width. The value of $\Delta t$ is a free parameter, which is to be determined from a scalegram analysis. The corresponding test functions are shown in Fig. 3.

There are three numerical criteria to determine the optimal value of the filter half-width $\Delta t$, namely, the r.m.s deviation of observations from the approximation $\sigma_0$, the r.m.s. accuracy of the approximation at times of observations $\sigma[x_C]$, the amplitude signal-to-noise ratio $S/N$. The test function $\sigma_0$ is nearly constant at $\Delta t \ll P$ (as the systematic differences of the observations from the approximation are negligible), as well as at $\Delta t \gg P$ (when the approximation asymptotically becomes a parabola). These nearly constant values ($\sigma_<$ and $\sigma_>$, respectively) may be used for an estimate of the characteristic semi-amplitude $R_\sigma = (2(\sigma_>^2 - \sigma_<^2))^{1/2}$ [18].

For this sample, minimal value $\sigma_< = 0.1785^m$ occurs at $\Delta t = 18^d$, so $R_\sigma = (2(0.9219^2 - 0.1785^2))^{1/2} = 1.279^m$. This value is definitely larger than that determined from the TP1 approximation ($R_{TP} = 1.13^m$) indicating systematic deviations from a pure sinusoid (either due to possible harmonics of the periodic signal, or to presence of aperiodic events like brightenings/weakenings or period variations).

We even propose a dimensionless parameter of sinusoidality $R^* = R_{TP}/R_\sigma$. This parameter is in a range from 0 to 1. For DY Per, it is equal to $R^* = R_{TP}/R_\sigma = 1.13^m / 1.279^m = 0.88$.

Another scalegram was proposed by [11]. It is based on the $\sigma_0$ scalegram and is proportional to $d\sigma_0^2/d(\lg \Delta t)$. Contrary to non-negative periodogram $S(f)$, $\Lambda(\Delta t)$ has more complicated shape even for a pure sine signal, including the intervals of negative values. It is shown If Fig. 3. The highest peak corresponds to $P_A=732.7^d$ and $R_A=0.971^m$. The second in height is a "hump" rather than a peak. It corresponds to $P_A=443.4^d$ and $R_A=0.545^m$. This seems to be not a "period", but a characteristic time scale of sharper events at the light curve.

**Approximations in Separate Intervals**

In the corresponding methods, the whole interval of observations is split into smaller intervals containing the extremum and parts of the nearby ascending and descending branches. Typically, the only information extracted from the approximation, is the moment of Minimum/Maximum (ToM, according to the terminology of the AAVSO [19]), i.e. "extremum". It is used for the *O-C* analysis [8,9,20]. Obviously, the simplest function with an extremum is a parabola (P2=polynomial of order 2). For distinctly asymmetric extrema, one may use a cubic polynomial (P3) [8]. Generally, the degree of the polynomial should be determined automatically [10] to determine the parameter with a best accuracy. This algorithm was implemented many times, using various computer languages [10,14,20-24]. For symmetrical extrema, the simplest improvement of the parabola, is the symmetrical polynomial [10,20,23].

Pulsating stars generally have asymmetric extrema. However, for rare noisy observations, the statistically optimal method may correspond to some symmetric function, as the number of parameters, which describe an asymmetry, vanish.

The general often disadvantage of the algebraic polynomial is the presence of the apparent waves in the approximation, which are similar the Gibbs phenomenon in the trigonometric polynomial approximations. To avoid such waves, [25,26] proposed an "asymptotic parabola" (AP) – the interval is split into three sub-intervals. Two straight lines ("asymptotes") are connected with a parabola, so the function and its first derivative are continuous. Contrary to polynomial splines, where the degree of the polynomial is generally constant, in the AP, the degree varies, as 1,2,1. Moreover, in AP, the number of subintervals is fixed to 3, but the borders are free parameters. The typical recommendation is to mark the intervals near extremum as wide as possible till the curve will have parabolic-like parts at the borders of the whole interval.

To improve the approximation for a wider interval (assuming they are symmetrical), two modifications were proposed: the parabolic spline (PS) [27] and the "Wall-supported" (WS) AP. Other WS algorithms are WSP (WS Parabola; effective for systems with transit eclipses) and WSL (WS Line; effective for systems with total eclipses) [28]. WS algorithms may be applied not only for typical eclipsing binary stars, but also for the systems with very different sizes of objects (either stars or exoplanets).

For the extrema without (nearly) flat parts, the interval may be split into two sub-intervals, where the border between them is the position of symmetry and thus of the extremum. Andronov [29,30] and Mikulasek [31] proposed special "shapes" (="templates" ="patterns") to approximate symmetric signals near extrema. However, for narrow intervals, these functions were converted to Taylor power series, and only 3 first terms are used:

Andronov et al. [32] tested 50+ different functions and range them according to a quality of the ToM determination.

These methods are implemented in the software MAVKA, which is still in progress. Totally there are 11 classes of functions used for the approximation

**Approximations using "Running Sines"**

The complete set of equations describing statistical properties of the running approximations using arbitrary basic and weight functions, was presented by Andronov [17]. Partially, it was implemented to the wavelet analysis [12,13] and was applied to pulsating stars e.g. by [33, 34]. However, the accuracy of the period determined from a short interval (e.g. in the wavelet analysis[35]) is much worth than the global approximations or using the $O$-$C$ analysis. So the algorithm of "Running Sines" (RS) [36] algorithm fills the gap between the global sine approximation and local wavelet fit.

Four parameters obtained using the RS approximation, are shown at the bottom part of Fig. 2. Smooth variations of the mean (over one period) brightness varies and corresponds to a "period" seen at the periodogram (Fig. 1), which is longer than the data, and thus is doubtful. However, the variations of $C_1$ are large: $11.22^m$–$12.69^m$. The semi-amplitude $R$ ranges from $0.44^m$ to $1.90^m$. This difference by a factor of ~4 is smaller than in another semi-regular variable RU And ($0.027^m$–$1.204^m$) [33]. The RS approximation varies from $10.49^m$ to $14.49^m$, whereas the AP approximation ($\Delta t=18^d$) shows different range ($10.68^m$–$16.12^m$). In both cases, it is caused by narrow asinusoidal faintings near JD 2452475 and JD 2456286, respectively.

The phase of the RS approximation was computed using the light elements (=ephemeris) obtained in the section "Periodogram Analysis":

Max. JD = $2454431.4(\pm2.5^d) + (794.6^d\pm1.0^d)\cdot E$ (1)

The zero cycle number is defined in such a way, that the initial epoch $T_0$ is the closest to a sample mean time $T_{mean}$, i.e. $|T_0 - T_{mean}| \leq 0.5P$, according to recommendations by [10]. This differs from a common definition of $T_0$ as the first ToM in a sample, so the cycle numbers start from zero towards positive integer numbers.

At a larger vertical scale, the dependence of phase on time is shown in Fig. 4. Besides low-amplitude waves caused by non-sinusoidal shape of the light curve, there are trends of different signs. This argues for a "switch" between the periods at this epoch. The AP approximation shows that the minimum of the test function (a sum of squares of the residuals) corresponds to a "zero-length" parabolic transition between the asymptotes, contrary, e.g. to another semi-regular star [37].

The light elements for these two parts of the light curve before and after the switch at JD $2454187\pm9^d$ are:

Max. JD = $2452847.4 (\pm3.0^d) + (851.1^d\pm4.1^d)\cdot E$ (2)

Max. JD = $2456020.7 (\pm4.1^d) + (780.5^d\pm2.7^d)\cdot E$ (3)

The statistical errors are formally small because of the large number of observations. The systematic deviations of the observations from the TP approximation are large, but the ~9.6% difference in the periods is statistically significant.

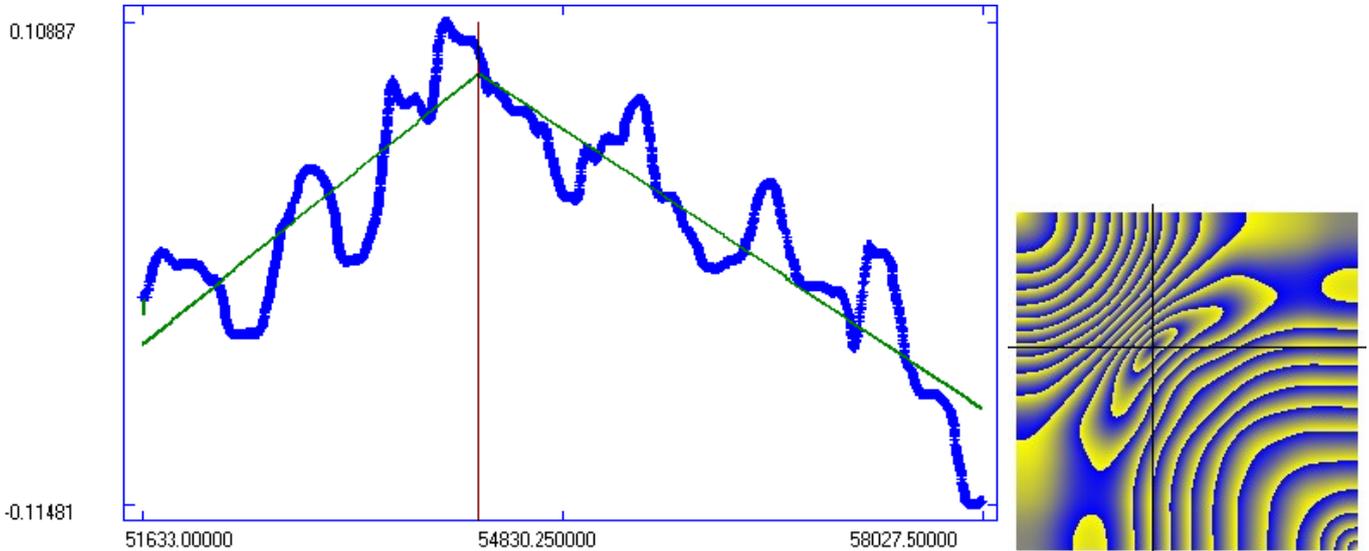
Fig. 4. The screenshots from the software MAVKA on the (left) dependence of phase of the RS ("Running Sine") [36] approximation of DY Per on time (JD-2400000) (blue) and its approximation by the AP ("Asymptotic Parabola"). The vertical line corresponds to the moment of "switching" between the periods at JD 2454187±9$^d$. The right figure represents the dependence of the test function on positions of the left and right borders of the inner parabola (see [28] for more details). For a better representation, the color of the pixel is not changed smoothly from minimum to maximum (as in common graphic representations), but has jumps to see isolines [28].

**Some Recommendations for Different Types**

In this paper, the methods are illustrated by an application to DY Per, the prototype of the class intermediate between SRb and RCrB. We compared approximations of three types: global, running and local.

For other stars, which are characterized by stable periodicity, one may recommend to use global trigonometric polynomial (TP) fits of statistically optimal order [10,38,39] (for pulsating and eclipsing (EB, EW subtypes) stars) or the "New Algol Variable" (NAV) [29,30,40,41] with special shapes (applicable not only for the EA, but also to EB and EW systems). These methods use a complete phase curve.

For shorter intervals near extremum, containing completely the ascending and descending branches, the approximations may vary from time-consuming "asymmetric hyperbolic secant" [20] and "log-normal-like" BSK [42] to faster methods with splitting an interval to two or three subintervals. These faster algorithms are implemented in the software MAVKA and were applied to eclipsing (e.g. [43-48]), symbiotic [49-51] and pulsating (e.g., [52-53]) stars.

For cataclysmic variables, sometimes the extremum is missing, but available the moment of crossing of the smoothing curve by an inverse approximation $t(m)$ instead of an usual $m(t)$ [54,55]. This may be done either in MCV, or in MAVKA. Other approximation (applied to intermediate polars) is a two-period model with (possibly) some harmonics [56-58]. This is available in MCV.

**Conclusions**

The net of complementary methods of data analysis should be used to study different types of variability.

**Acknowledgements.**

This work was initiated by Dr. Bogdan Wszołek, who makes a great contribution to public outreach and popularization of astronomy via the international society "Astronomia Nova".

We thank the French Association of Variable Stars Observers AFOEV (Association Francaise des Observateurs d'Etoiles Variables, http://cdsarc.u-strasbg.fr/afoev) for a huge number of observations of variable stars made by amateurs and available on-line, particularly, DY Per. This work is a part of the "Stellar Bell" [58] part of the "Inter-Longitude Astronomy" [60,61] international project, as well as of the `'Ukrainian Virtual Observatory" [62,63] and "AstroInformatics" [64].